\documentclass[aip,pop,reprint]{revtex4-1}
\usepackage[pdftex]{graphicx}
\usepackage{graphicx}
\usepackage{subfigure}
\usepackage{amsmath}
\usepackage{amsfonts}
\usepackage{mathtools}
\usepackage{geometry}

\begin{document}

\title{The Madison plasma dynamo experiment: a facility for studying laboratory plasma astrophysics}

\author{C. M. Cooper}
\affiliation{Department of Physics, University of Wisconsin, Madison WI 53706}
\affiliation{Center for Magnetic Self Organization, University of Wisconsin, Madison WI 53706}
\author{J. Wallace}
\affiliation{Department of Physics, University of Wisconsin, Madison WI 53706}
\author{M. Brookhart}
\affiliation{Department of Physics, University of Wisconsin, Madison WI 53706}
\affiliation{Center for Magnetic Self Organization, University of Wisconsin, Madison WI 53706}
\author{M. Clark}
\affiliation{Department of Physics, University of Wisconsin, Madison WI 53706}
\author{C. Collins}
\affiliation{Department of Physics, University of Wisconsin, Madison WI 53706}
\affiliation{Center for Magnetic Self Organization, University of Wisconsin, Madison WI 53706}
\author{W. X. Ding}
\affiliation{Department of Physics and Astronomy, University of California, Los Angeles, Los Angeles, CA 90024}
\author{K. Flanagan}
\affiliation{Department of Physics, University of Wisconsin, Madison WI 53706}
\author{I. Khalzov}
\affiliation{Department of Physics, University of Wisconsin, Madison WI 53706}
\affiliation{Center for Magnetic Self Organization, University of Wisconsin, Madison WI 53706}
\author{Y. Li}
\affiliation{Department of Physics, University of Wisconsin, Madison WI 53706}
\author{J. Milhone}
\affiliation{Department of Physics, University of Wisconsin, Madison WI 53706}
\affiliation{Center for Magnetic Self Organization, University of Wisconsin, Madison WI 53706}
\author{M. Nornberg}
\affiliation{Department of Physics, University of Wisconsin, Madison WI 53706}
\affiliation{Center for Magnetic Self Organization, University of Wisconsin, Madison WI 53706}
\author{P. Nonn}
\affiliation{Department of Physics, University of Wisconsin, Madison WI 53706}
\author{D. Weisberg}
\affiliation{Department of Physics, University of Wisconsin, Madison WI 53706}
\affiliation{Center for Magnetic Self Organization, University of Wisconsin, Madison WI 53706}
\author{D. G. Whyte}
\affiliation{Plasma Science and Fusion Center, Massachusetts Institute of Technology, Cambridge MA 02139}
\author{E. Zweibel}
\affiliation{Department of Physics, University of Wisconsin, Madison WI 53706}
\affiliation{Department of Astronomy, University of Wisconsin, Madison WI 53706}
\affiliation{Center for Magnetic Self Organization, University of Wisconsin, Madison WI 53706}
\author{C. B. Forest}
\affiliation{Department of Physics, University of Wisconsin, Madison WI 53706}
\affiliation{Center for Magnetic Self Organization, University of Wisconsin, Madison WI 53706}
\date{\today}

\begin{abstract}
The Madison plasma dynamo experiment (MPDX) is a novel, versatile, basic plasma research device designed to investigate flow driven magnetohydrodynamic (MHD) instabilities and other high-$\beta$ phenomena with astrophysically relevant parameters.   A 3 m diameter vacuum vessel is lined with 36 rings of alternately oriented 4000 G samarium cobalt magnets which create an axisymmetric multicusp that contains $\sim$14 m$^{3}$ of nearly magnetic field free plasma that is well confined and highly ionized $(>50\%)$.  At present, 8 lanthanum hexaboride (LaB$_6$) cathodes and 10 molybdenum anodes are inserted into the vessel and biased up to 500 V, drawing 40 A each cathode, ionizing a low pressure Ar or He fill gas and heating it.  Up to 100 kW of electron cyclotron heating (ECH) power is planned for additional electron heating.  The LaB$_6$ cathodes are positioned in the magnetized edge to drive toroidal rotation through ${\bf J}\times{\bf B}$ torques that propagate into the unmagnetized core plasma.  Dynamo studies on MPDX require a high magnetic Reynolds number $Rm > 1000$, and an adjustable fluid Reynolds number $10< Re <1000$, in the regime where the kinetic energy of the flow exceeds the magnetic energy ($M_A^2=($v$/$v$_A)^2 > 1$).  Initial results from MPDX are presented along with a 0-dimensional power and particle balance model to predict the viscosity and resistivity to achieve dynamo action.
\end{abstract}


\maketitle

\section{Introduction}
The Madison plasma dynamo experiment (MPDX) is designed to create a steady-state, hot, weakly magnetized, flowing plasma in which the kinetic energy of the flow drives magnetohydrodynamic (MHD) instabilities.  Such a device accesses a new regime relevant to astrophysical applications and never before achieved in a laboratory\cite{spence09_apj}.  This plasma is well-suited for studying astrophysical phenomena such as the dynamo process \cite{Ossendrijver2003Solar}, the feasibility of which is the topic of several recent publications \cite{Khalzov2012, Khalzov2012a,khalzov2013} and is investigated in section \ref{subsect_scenarios}.  The magneto-rotational instability (MRI)\cite{Balbus1998Instability} could also be investigated using the tunable boundary driven flow or a supplemental central spinning post in conjunction with an external Helmholtz coil. In addition, many basic plasma physics studying interactions and organization of flows and magnetic fields can be performed.

Previous and ongoing dynamo experiments \cite{Gailitis2002Colloquium} use flowing liquid metals which can be stirred mechanically and confined with a simple vessel.  These interesting experiments suffer from several limitations which can be avoided by using plasmas.  First, the magnetic Reynolds number achieved in liquid metal experiments, which governs the transition to a dynamo, is too low.  The magnetic Reynolds number $Rm=$v$L/\eta$ (where $L$ is the characteristic size, and $\eta$ the resistivity and v is the flow velocity) is the ratio of magnetic field advection by the flow to magnetic field diffusion from the resistivity.  In liquid metal experiments, $Rm$ is typically less than $Rm_{crit}$ necessary for dynamo and MRI excitation.  In addition, the fluid Reynolds number, which is the ratio of the momentum advection by the flow to the momentum diffusion by the viscosity $Re=$v$L/\nu$ (where $\nu$ is the viscosity), cannot be varied independently of $Rm$ in liquid metals.  The ratio of these quantities, the magnetic Prandtl number $Pm=Rm/Re=\nu/\eta$ is a fixed, very small value $\sim 10^{-5}$ in liquid metal; thus, the flows required to exceed $Rm_{crit}$ are always turbulent.  In plasmas, by contrast, $Rm$ can be varied independently of $Re$ by controlling the plasma resistivity and viscosity by changing the electron temperature, the plasma density, and the ion species.  $Pm \sim 1$ is achievable in MPDX making these plasmas are excellent candidates for dynamo action and comparison to simulations.  In addition, laminar dynamo action ($Rm>Rm_{crit}, Re<\sim1000$) can be studied in the absence of fluid turbulence which can suppress it\cite{Rahbarnia2012}.

To experimentally test theories about the dynamo mechanism in a plasma, the device operates in the {\it flow-dominated} regime as opposed to the {\it magnetically-dominated} regime.  Such an experiment requires not only that $Rm$ be large, but also that the flow energy dominates over the magnetic energy.  This is quantified by the Alfv\'{e}n Mach number squared $M_A^2=($v$/$v$_A)^2$ which is the ratio of the inertial force to the magnetic force, where v$_A=B/\sqrt{\mu_on_im_i}$ is the Alfv\'{e}n velocity.  When $M_A \gg 1$ and $Rm\gg 1$ flows can stretch and amplify the magnetic field.  The feasibility of creating such a plasma in the lab has recently been demonstrated in a novel multi cusp confined plasma with electrostatic stirring \cite{katz2012_rsi} where unmagnetized  plasmas with flows of  $\sim 10$ km/s have been created \cite{collins2012}.  For MPDX class plasmas, a convenient set of formula are
\begin{eqnarray}
 Re & = & 7.8 \; \text{v}_{km/s} L_m n_{i,10^{18} m^{-3}}Z^4\sqrt{\mu}/T_{i,eV}^{5/2} \\
 Rm & =& 1.6 \; T_{e,eV}^{3/2} \text{v}_{km/s} L_m/Z \\
 M_A & = &0.46 \;   \text{v}_{km/s} \sqrt{ \mu n_{i,10^{18} m^{-3}} } / B_g,
 \label{eq_ReRm}
\end{eqnarray}
where the Braginskii viscosity and the Spitzer resistivity for unmagnetized plasma are used for $\nu$ and $\eta$ respectively, $n_{i,10^{18}}$ is the ion density in $10^{18}$/m$^3$, $\mu$ is the atomic mass in amu, $Z$ is the net charge of the ions, $T_{e,eV}$ and $T_{i,eV}$ are the electron and ion temperatures respectively in eV,  $L_m$ is the characteristic size in meters, v$_{km/s}$ is the plasma velocity in units of km/s and $B_g$ is a characteristic strength of the background magnetic field in Gauss.  The plasma parameters and relevant dimensionless parameters are adjusted by controlling the neutral fill pressure, input power, gas species, and torques imposed on the plasma edge. 

\begin{figure}
 \includegraphics[width=1.0\columnwidth]{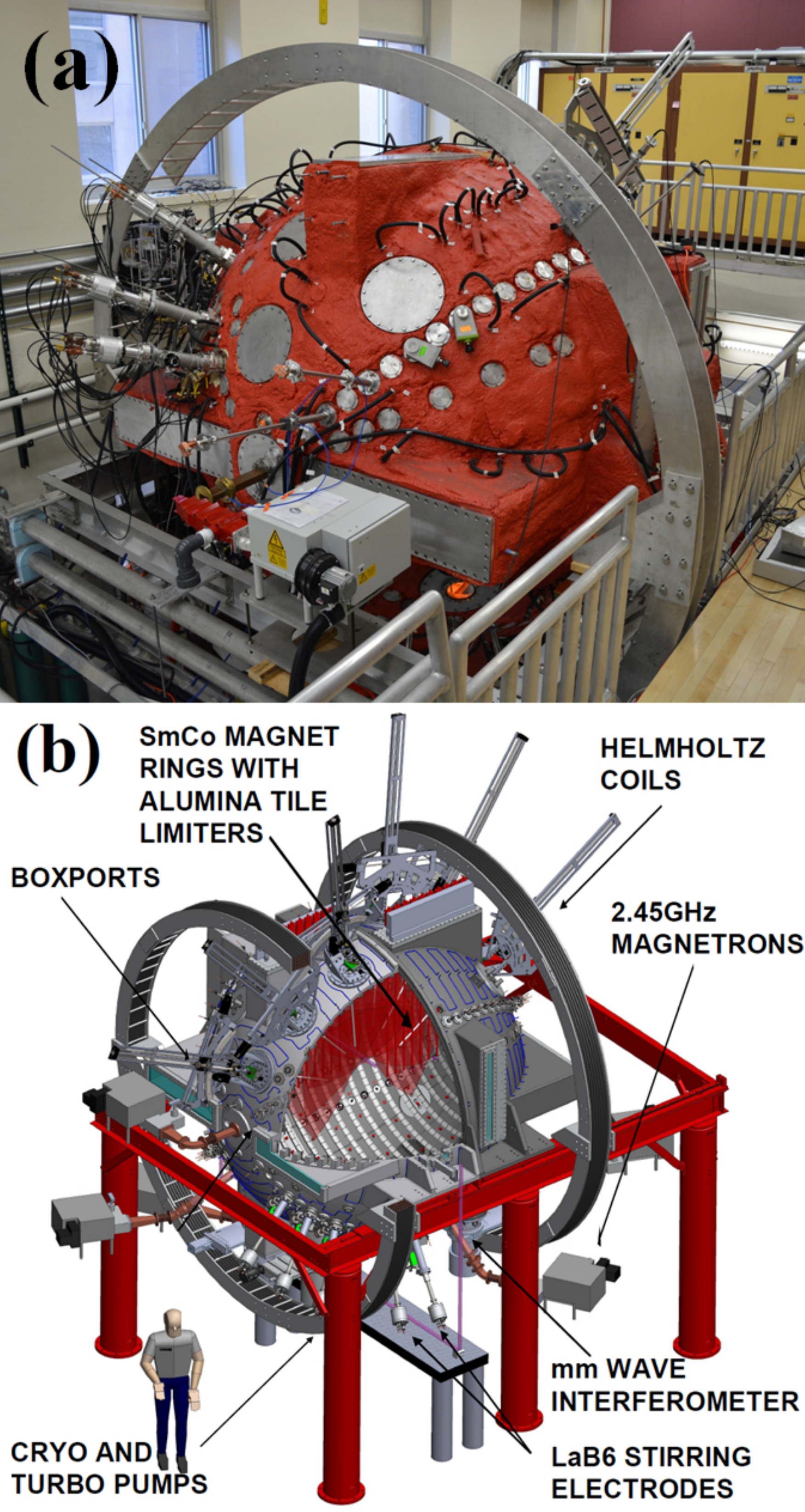}
 \label{fig:annotated_MPDX_geometry}
 \caption{(a) a picture of MPDX and (b) the various current and planned subsystems labeled.}
\end{figure}

The MPDX uses several well-established techniques to confine and stir a hot, unmagnetized plasma.  First, an axisymmetric multipole magnetic ring cusp built from alternating rings of individual permanent magnets creates a strong magnetic field localized to the edge, forming a magnetic bucket to confine the plasma.  Second, biased hot cathodes draw current to induce an ${\bf J} \times {\bf B}$ rotation to stir the edge of the plasma.  The flow is driven at the boundary and momentum is viscously transported throughout the unmagnetized region.  By controlling the spatial profile of these boundary-driven flows, arbitrary large scale helical flows can be established (Fig. \ref{fig:annotated_MPDX_geometry}).

The paper is organized as follows: Section \ref{sect_desc} provides a description of the device, its construction and initial operation.  Section \ref{sect_LaB6} describes the LaB$_6$ cathodes which are used for producing plasma and controlling the boundary flow.  Section \ref{sect_diagnostics} describes the current state of diagnostics on MPDX.  Section \ref{sect_results} presents initial measurements of plasma parameters in the MPDX operation that provide the basis for scaling to higher power.  Section \ref{sect_sim} outlines simulations predicting machine confinement and operation as well as providing a plan for achieving several types of dynamo action in MPDX.

\section{Description of the MPDX}
\label{sect_desc}
New parameter regimes in laboratory plasmas can be reached due to recent breakthroughs in plasma physics technologies.  This section describes the design of the facility, vacuum chamber, magnets, and plasma sources. 

\subsection{Vacuum Chamber}
\begin{figure}
 \includegraphics[width=1.0\columnwidth]{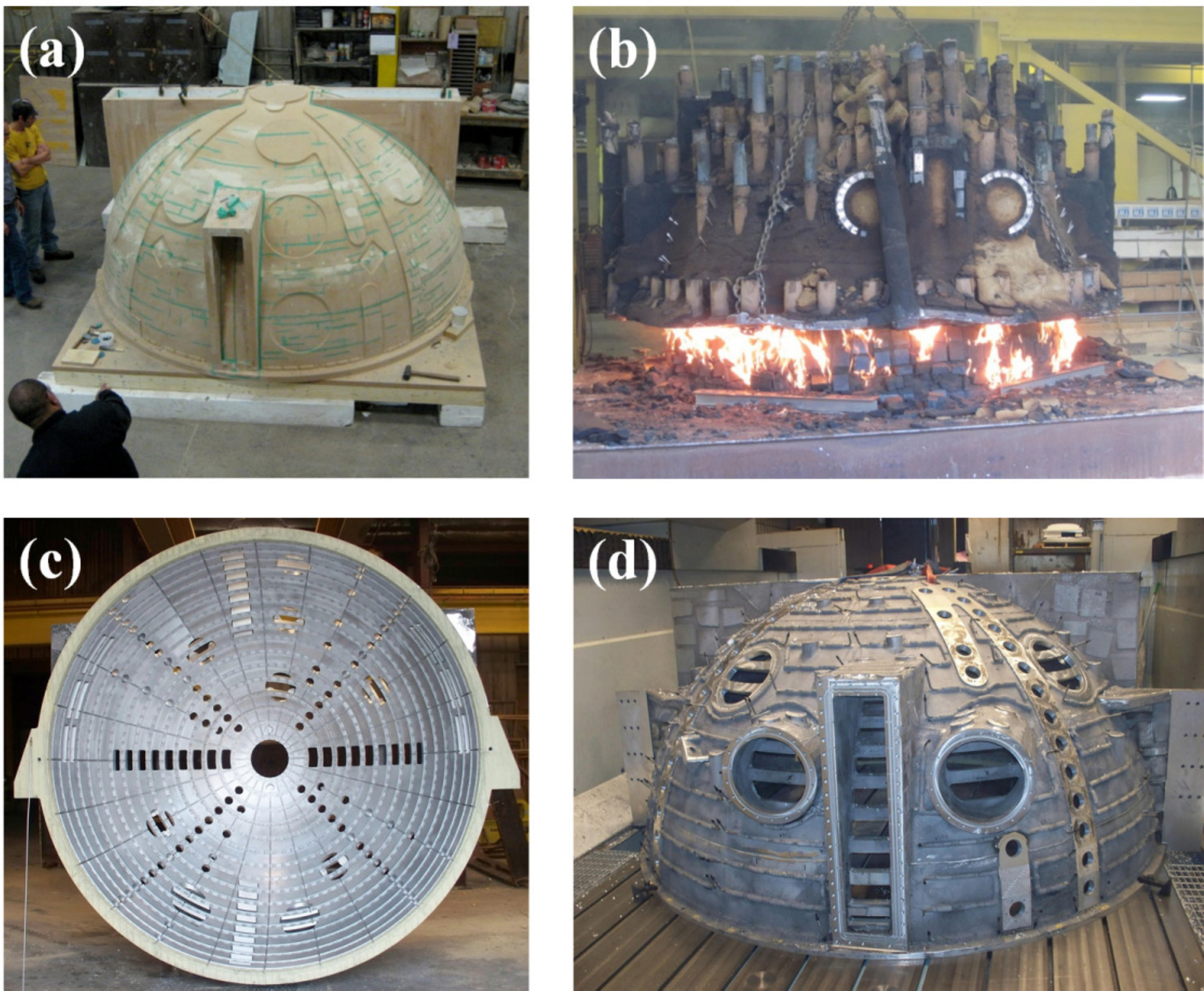}
 \caption{(a) The positive of the chamber used to make the mold.  (b) The casting process for the vessel in the mold.  (c) The fully machined interior of one hemisphere of the vacuum vessel. Precision located rings with planar facets and tapped holes for each magnet provide for the cusp field.  (d) The exterior of the vacuum vessel showing ports, magnet bridging, and embedded vessel cooling.}
 \label{fig:pics_of_casting}
\end{figure}

The experimental vacuum chamber is comprised of two 3 m diameter hemispherical shells joined by equatorial flanges to create a spherical vessel.  One hemisphere is mounted on a linear stage to provide easy opening and access to the chamber interior.  The hemispherical shells are made of 0.03 m thick cast A356 aluminum (see Fig. \ref{fig:pics_of_casting}).  The wall thickness is determined by structural integrity and the need to provide sufficient material to machine flanges and other features into the walls.

Each hemisphere is cast as a single piece by Portage Casting and Molding, Inc. which provides many benefits over a traditional welded vacuum vessel.  Casting not only eliminates the need for costly vacuum welds, but has allowed for a more complex and precise design at a greatly reduced cost.  The design process resembled ``3D Printing:'' Solidworks\cite{Solidworks} computer aided design software was used by the UW engineering team to design the machine.  The design was sent to the aluminum casting firm where a 5 axis milling machine was used to construct a positive (i.e. pattern) for the mold. Sand molds were made, and molten aluminum was cast into the molds.  Stainless steel tubular cooling lines are cast directly into the walls providing a low thermal impedance to remove heat from the experiment.  Fig. \ref{fig:pics_of_casting} show the construction.

The vessel features six identical boxports (0.2 m x 1.0 m) on each hemisphere to provide large views for optical diagnostics.  Each hemisphere has seven larger 0.35 m circular ports for probes and Radio-Frequency (RF) heating sources.  Each pole has a 0.35 m port for RF feedthroughs and to provide axial access.  An additional 182 ports 0.075 m in diameter are located on the sphere between magnet rings for insertion of plasma stirring electrodes, diagnostics or electrical connections.  The chamber interior is plasma spray coated with a 0.00025 m thick alumina film.  The large diameter seals (including the seals between the two hemispheres) are differentially pumped double O-ring seals.  Two 2,000 L/m turbo pumps and two 4,000 L/m cryo pumps routinely achieve $5\times10^{-7}$ Torr vacuum base pressure, similar to stainless steel vessels sealed with O-rings.  The experimental operating pressure with gas fill is $< 5\times10^{-4}$ Torr.
\subsection{Magnets}

Eighteen faceted rings are precisely machined into each hemisphere interior and used for mounting the $\sim$ 3000 0.038 m $\times$ 0.025 m $\times$ 0.05 m samarium cobalt (SmCo) magnets.  Unlike neodymium (NdFeB) magnets, SmCo magnets do not corrode in the presence of hydrogen gas.  Each ring of magnets alternates in polarity, creating a 36 pole, axisymmetric magnetic field cusp for plasma confinement.  The magnets are in good thermal contact with the water cooled vacuum vessel and operate at temperatures up to 40$^o$C, far below their curie temperature of 350$^o$C.  The magnets are covered by Kapton film and thin interlocking alumina ceramic tiles that protect and electrically isolate them.

\begin{figure}
 \includegraphics[width=1.0\columnwidth]{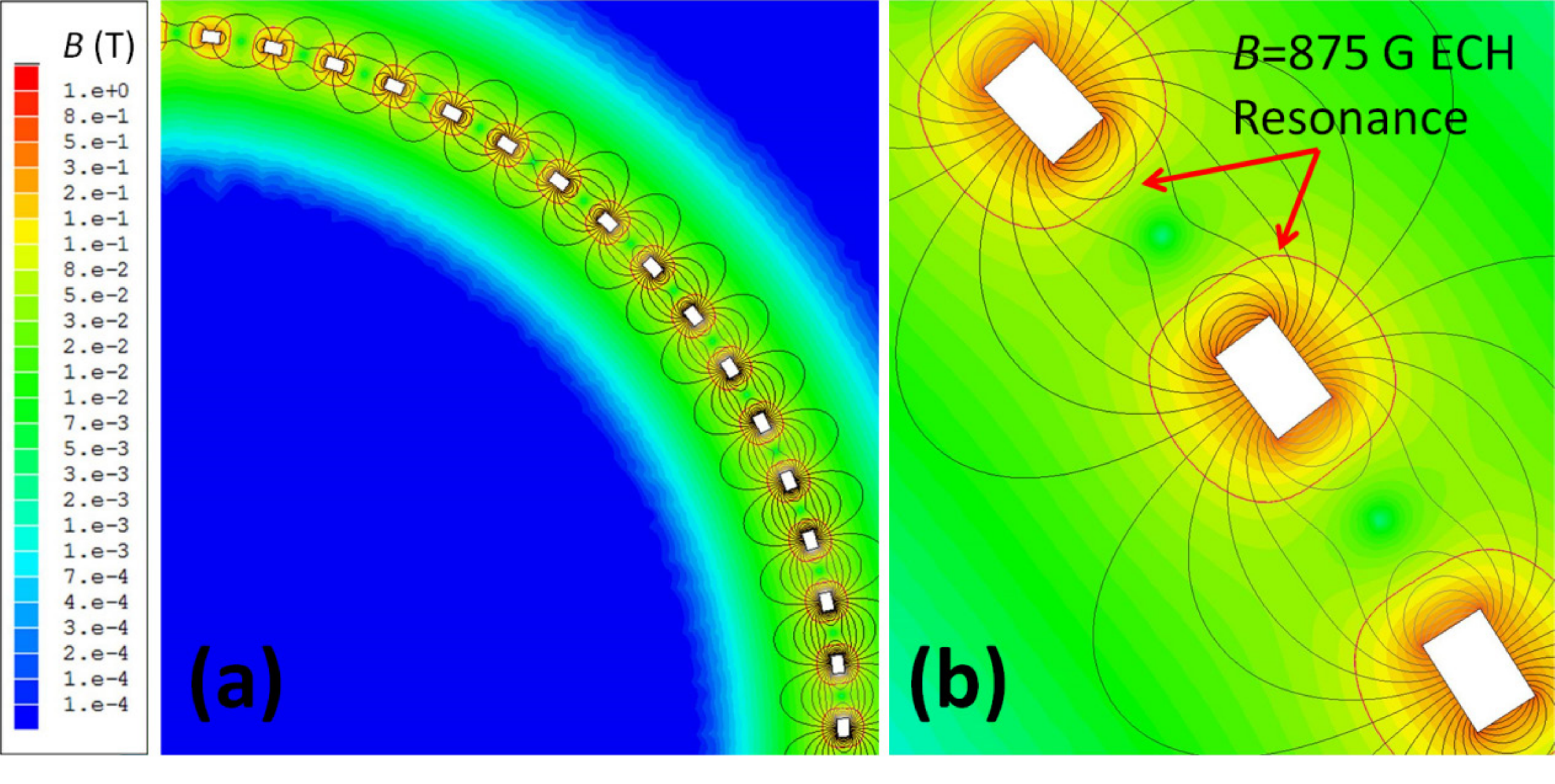}
 \caption{(a) Poloidal magnetic flux contours for the MPDX (magnetic field lines) and color contours of the magnetic field strength as computed by the MaxwellSV finite element modeling package and (b) a closeup of the cusp.  The magnetic field is generated by axisymmetric rows of 4000 G SmCo magnets with alternating polarity.  The central dark blue contour shows the region where the magnet-produced field is less than 0.2 gauss and the red contour shows the 875 G ECH resonance surface.}
 \label{fig_maxwell} 
\end{figure}

The primary role of the magnets is to confine the large unmagnetized spherical volume of plasma for basic plasma experiments.  The result from a Maxwell SV finite element modeling of the magnetic field used to design the magnet geometry is shown in Fig. \ref{fig_maxwell}.  The magnetic field is $<1$ G 0.25 m away from the magnet face.  Residue marks on the tiles appeared after 10 hrs of integrated ``plasma on'' time and indicate the footprint of the plasma contact with the tiles.  They are 0.0008 m wide, in good agreement with the width predicted by theory for our plasma parameters\cite{hershkowitz75_prl}.  The magnet ring spacing in the MPDX is 0.11 m which implies that loss area is less than  1\% of the surface area of the vessel.  The MPDX confines $\sim$ 14 m$^3$ of plasma with an convective plasma loss surface area of $\sim$ 0.2 m$^2$. Particle and energy confinement pertaining to achievable parameters in MPDX are further investigated in section \ref{sect_sim}.  

The cusp field also plays an important role in providing a surface for a 2.45 GHz electron-cyclotron resonance (where $B=875$ G, approximately 0.05 m in front of the magnets as seen in Fig. \ref{fig_maxwell}).  Five magnetrons manufactured by CoberMuegge, LLC, are being installed to provide up to 100 kW continuous power for ionizing and heating the plasma.  Simple unterminated waveguide is used for antennas in either O-mode or X-mode polarization relative to the direction of the cusp magnetic field.  Such sources are routinely used for heating overdense plasmas ($n_e$ greater than the critical density $7\times 10^{16}$ m$^{-3}$ for 2.45 GHz).  This is speculated to be either be due to mode conversion to the electron Bernstein wave or by parallel propagation from the high field side of R-waves (whistler waves), neither of which suffers from the cutoff.  Additionally, the magnetic fields are necessary for generating flows in conjunction with the LaB$_6$ cathodes and molybdenum anodes.

An external Helmholtz coil is now being installed for introducing a seed magnetic field and to probe the interactions between the driven flows and a scalable background magnetic field.

\subsection{LaB$_6$ Stirring Electrodes}
\label{sect_LaB6}

Lanthanum Hexaboride has long been used as a high temperature ($T_e<30$ eV), high density ($n_e<2\times 10^{19}$ m$^{-3}$) plasma source due to its high input power density and robust construction \cite{ono87,schmitz1990, goebel2005, cooper2010}.  The ability to stir the MPDX plasma through LaB$_6$ cusp biasing is possible because LaB$_6$ can withstand high bias voltages ($V_d<500$ V).  

\begin{figure}
 \includegraphics[width=1.0\columnwidth]{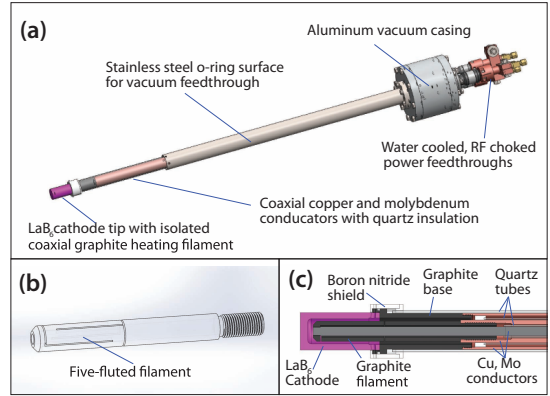}
 \caption{LaB$_6$ stirring electrode design.  The entire (1.1 meter long) electrode is shown in (a), with detailed views of (b) the graphite heater filament and (c) the LaB$_6$ tip.  Not shown: the linear vacuum feedthrough and garage / gate-valve assembly.}
 \label{fig:cathode}
\end{figure}

The LaB$_6$ stirring electrode design developed for MPDX is shown in Fig. \ref{fig:cathode}.  The cylindrical LaB$_6$ cathode (measuring 0.056 m in length, 0.0264 m in diameter, with 0.003 m thick walls) is radiatively heated up to 1600 degrees Celsius by an internal graphite heater filament.  This filament is precision machined with a five-fluted current path and threaded ends to ensure tight electrical connections.  To prevent arcing, the heater filament is electrically isolated from the AC heater power supply with an isolation transformer and does not touch the LaB$_6$ cathode.  A stainless steel tube allows for a sliding vacuum seal, a press-fitted plastic bushing provides structural support for the 12 kg electrode, and a large stainless steel ``garage'' allows the cathode to be completely retracted behind a gate-valve for periodic maintenance.  The conductors for the heater circuit as well as the cathode discharge circuit are cylindrical copper tubes with a center molybdenum rod, coaxially nested inside the stainless steel tube and isolated from each other and the surrounding plasma by quartz tubing.  

In anticipation of the large amounts of RF power provided by MPDX's magnetrons, the three electrical feedthroughs per electrode are each paired with a copper clamp that fulfills three essential functions: electrical connection to power supply cables, water cooling to the cylindrical conductor bases, and transmission line choking of RF waves.      

\begin{figure}
 \includegraphics[width=1.0\columnwidth]{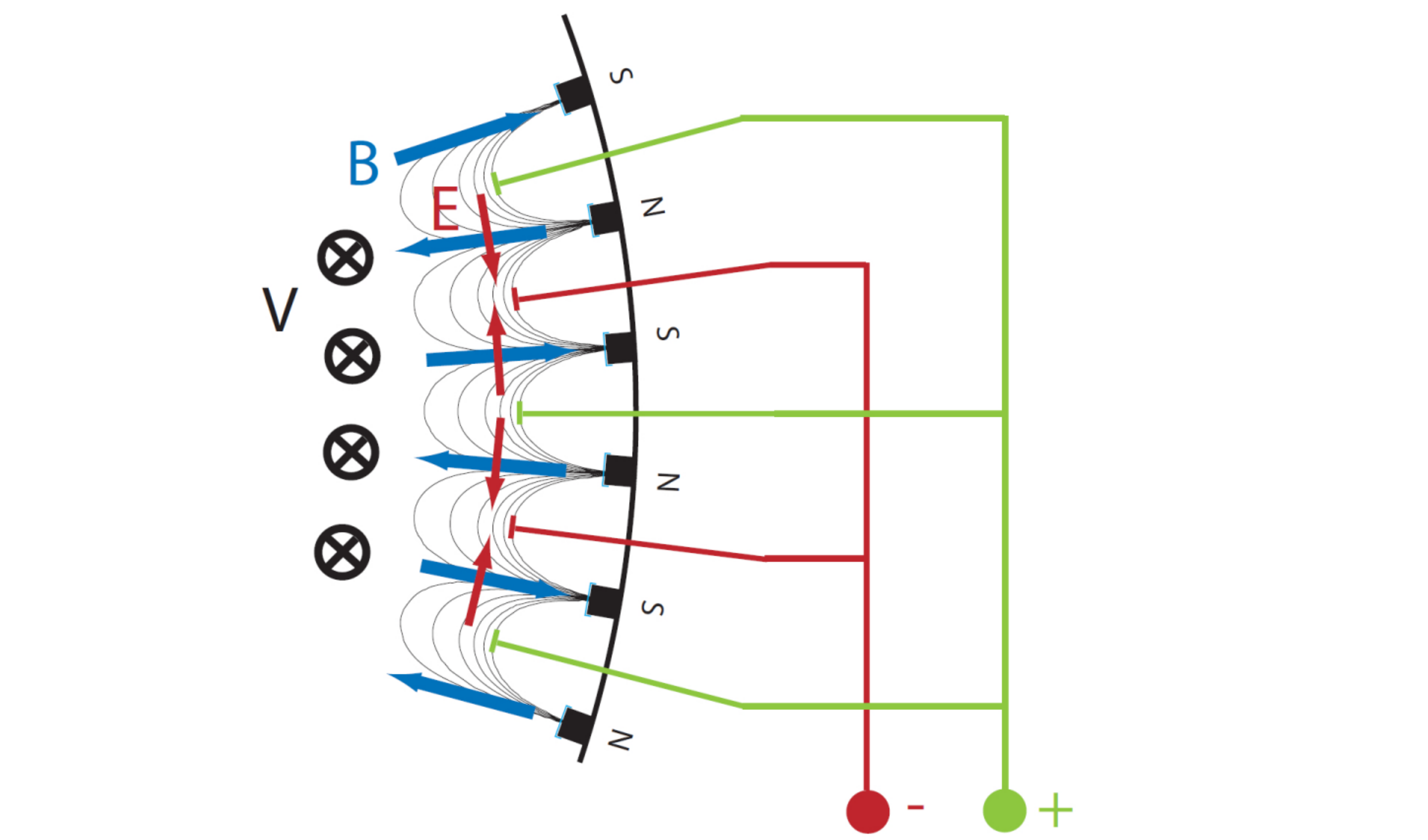}
 \caption{A cartoon of the flow drive mechanism in MPDX using biased anode-cathode pairs in the cusp field.  The torque is proportional to the anode-cathode bias voltage, current and their location in the cusp which determines $B$}
 \label{fig_flow} 
\end{figure}

Each LaB$_6$ stirring electrode is paired with one or more Molybdenum anodes poloidally displaced one cusp.  The current path between cathode and anode must cross a permanent magnet ring and produces ${\bf J}\times{\bf B}$ torque in the toroidal direction, as shown in Fig. \ref{fig_flow}.  Azimuthal particle drifts rapidly symmetrize the potential\cite{collins2012}, creating a virtual cathode connected to the actual cathode, but toroidally symmetric.  This edge-applied torque couples to the bulk plasma via ion-ion viscosity.  Initial experiments using these electrodes have injected up to 40 A of current at 500 V.  Each anode-cathode potential will eventually be independently set, creating a customizable poloidal profile of toroidal velocity.

\subsection{Diagnostics}
\label{sect_diagnostics}

The MPDX plasma is diagnosed by physical probes capable of making in situ measurements of local plasma parameters.  These probes can all be pumped down and inserted into the machine via vacuum gate valves without interrupting operation.  The plasma temperature and density are low enough for long probe lifetimes.  The surface area of the probes and probe shafts is much smaller than the loss area on the cusp, and individual probes do not perturb the plasma.

A swept voltage bias is applied to the tip of planar Langmuir probes to create an ``I-V'' curve.  This curve is analyzed to determine $T_e$, $n_e$, $\Phi_p$, and $\Phi_f$ using standard Langmuir probe techniques \cite{Hutchinson2002}.  Plasma flows inside MPDX are measured using Mach probes.  The probe tips are biased several times $T_e$ below the plasma potential to collect ion saturation current and the Mach number and absolute speed can be measured using standard techniques \cite{Hutchinson2002}.  A triple probe is used to provide an additional measurement of $T_e$ and $\Phi_f$.

A phase locked set of 1 mm sources acquired from Virginia Diodes, along with two mixers and an Analog Devices phase detector provide a single channel interferometric measurement of the line-integral density.  The 1 mm wavelength is optimal for normal MPDX densities of $\sim10^{17}-10^{18}$ m$^{-3}$ when integrated along a 6 m pathlength twice the length of the MPDX plasma. These densities correspond to phase shifts from $\sim\frac{\pi}{2}$ to $4\pi$, which are easily measurable.  The upper bound for a measurable density is $n_{e}\sim10^{21}$ m$^{-3}$ and the minimum detectable density set by the digitizer noise levels is $\sim10^{15}$ m$^{-3}$, a sufficient resolution for MPDX densities.  An example calibrated interferometer time trace is found in Fig. \ref{fig_3secshot}.

A bolometer made from a LiTa crystal is installed on the MPDX to measure the total radiated power.  This includes light radiated from the neutral particles and ions due to collisions with electrons, as well as the incident kinetic energy of any superthermal neutral gas created through elastic and charge exchange collisions.  The only power loss rate not included is the ionization power losses and the losses to the wall.  The bolometer is calibrated with a laser over a scaling of incident power.  
    
Magnetic diagnostics will consist of an array of Hall probes mounted on the MPDX surface to measure the multipole moment of the slowly varying magnetic fields.  Rogowski coils on probes inside the plasma will be calibrated and integrated\cite{Everson2009} to measure the weak, fluctuating magnetic fields.

A two dimensional movable probe drive is mounted on the MDPX vessel with a gate valve and a rotatable vacuum ball valve.  Two stepper motors mounted perpendicular to each other can change the position of the probe inside the machine, allowing the probe to trace out a plane of data as seen in Fig. \ref{fig_radprofile}.  The probe drive can be used with many types of probes in MPDX and is very important for studying 2D systems in MPDX.

The data collected by the diagnostics are digitized by a 500 kHz 96-channel ACQ-196 module from D-TACQ Solutions Ltd, recorded by a computer and stored for analysis.  Information about the MPDX such as the machine conditions, average plasma parameters, and probe positions are recorded in a searchable SQL database.  Probe movement and discharge parameters are controlled by a Labview interface via a Compact RIO.  The machine can be run in a pulsed operation or in a steady-state mode.  The MPDX has created over 3000 discharges and nearly 3 hours of ``plasma on'' time total.  Plasma discharges are fully automated and create 3-5 s plasma discharges every 45 s around the clock.

\section{Initial MPDX Results}
\label{sect_results}

The first major result from MPDX is exhibiting confinement of an unmagnetized plasma using the edge localized magnetic multipole cusp.  Eight LaB$_6$ cathodes ($\theta=30^o,40^o,50^o,60^o,125^o,135^o,145^o,155^o,$ $\phi=225^o$) and eight anodes displaced one cusp poloidially and $90^o$ toroidally ($\theta=35^o,45^o,55^o,65^o,120^o,130^o,140^o,150^o,$ $\phi=135^o$), were inserted radially into the unmagnetized region ($r$=1.25 m).   The cathodes were biased 200-500 V with respect to the anodes, with each cathode drawing 5-40 A for a discharge time of 3-10 s in order to characterize the breakdown and confinement in the MPDX.  Electron heating was provided exclusively by the Ohmic heating of the anode-cathode circuit.  

\begin{figure}
 \includegraphics[width=1.0\columnwidth]{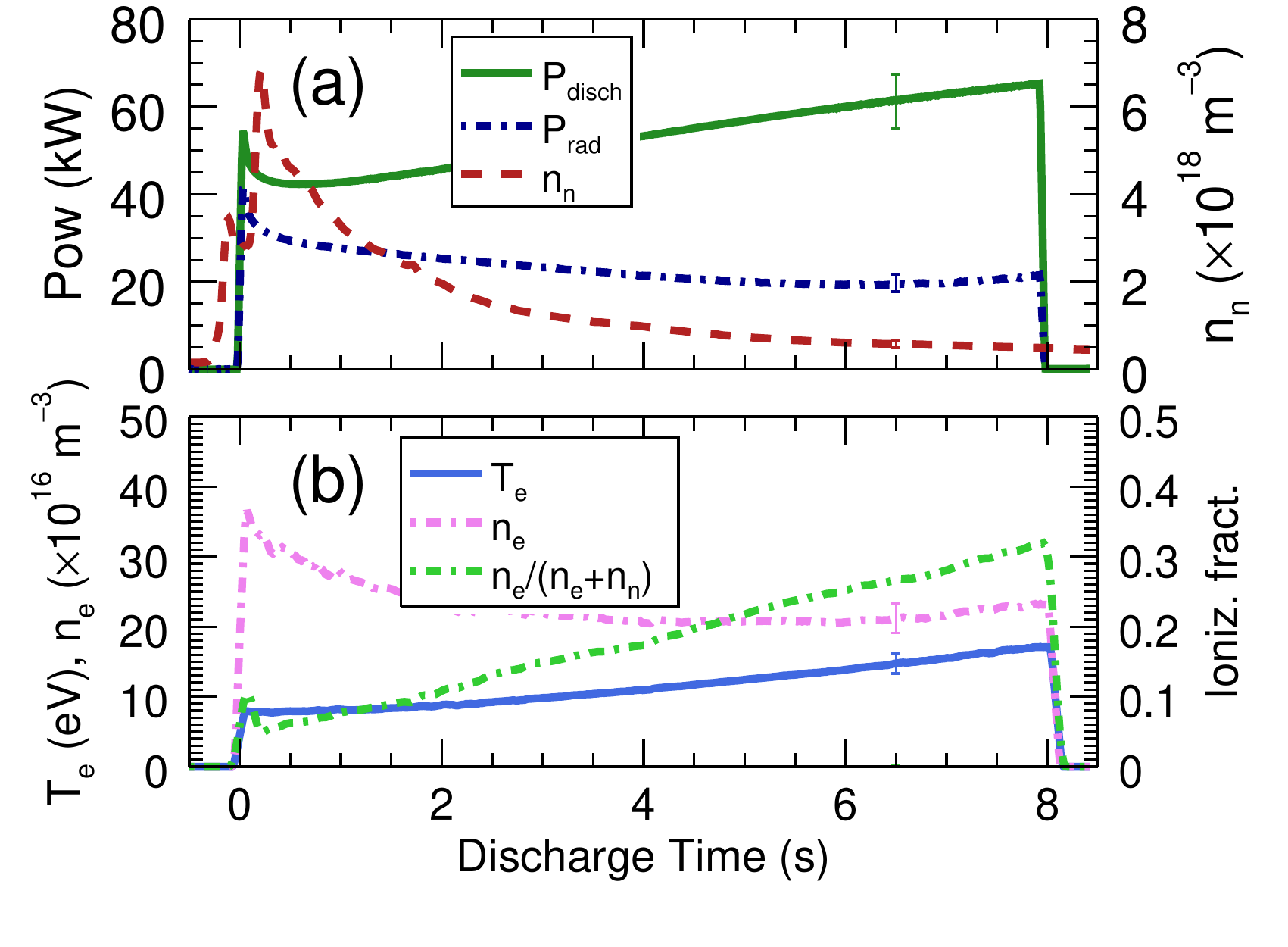}
 \caption{(a) Time trace of the gas puff, the input power, and the radiated power from the bolometer (b) $T_e$ measured by a Langmuir probe, $n_e$ measured by an interferometer normalized to a path length measured by probes, and the calculated ionization fraction $n_e/(n_n+n_e)$.}
 \label{fig_3secshot}
\end{figure}

The machine can create steady state plasma discharges with a constant input power and fill pressure where plasma parameters stay constant over the length of the shot.  However, the most interesting plasma regimes come where the neutral gas is puffed in, over a 20 ms time, 0.5 s before discharge, achieving a neutral fill pressure high enough to achieve breakdown that is then pumped out to achieve high ionization fractions and electron temperatures.  An example time trace for an eight second discharge at 400 V drawing a total of 150 A for a helium gas puff is shown in Fig. \ref{fig_3secshot}.  A Langmuir probe was inserted radially into the unmagnetized region ($r$=1.12 m), $\theta=110^o, \phi=45^o$ and continuously swept -80 V to + 20 V at 100 Hz to determine $T_e(t)$ in the core of MPDX.  The interferometer was directed through a box port, reflected off the far wall and collected on the same side.  The interferometer was normalized to the path length though the plasma as measured by probes to determine $n_e(t)$.  The radiated power from the plasma from the bolometer was integrated over the surface area of the machine to yield the total radiated power for the experiment.  The shot to shot reproducibility is comparable to the error in this measurement ($\delta T_e/T_e\sim 5\%$), which enables ensembeled averaging of datasets over multiple shots and with the probe at different locations.

The neutral gas density decreases as the gas is pumped out over the course of the shot.  With fewer neutral particles to ionize, the plasma density drops as well.  With fewer losses to ionization, the walls, and collisions with the neutral gas and ions, the electrons heat up with the increase in input power.  This is manifested in the ionization percentage, which increases from 5\% to 35\% over the course of the shot.  However, this is the core $n_e$ compared to $n_n$ measured at the wall.  In this plasma with $n_e\sim 2.5\times 10^{17}$ m$^{-3}$ and $T_e\sim 15$ eV, the neutral mean free path for ionization in the reference frame of the neutral particle (the neutral penetration depth) is 0.75 m, meaning the center of the plasma (two e-foldings in radius) is nearly fully ionized.  This will be investigated with passive spectroscopy.  In addition, the radiated power measured by the bolometer drops even as the total input power rises.  Using the below measured driven edge rotation speed of v$_\phi=$2.3 km/s in a similar discharge yields a local $Rm=320$, an important benchmark for studying the dynamo.
 
\begin{figure}
 \includegraphics[width=1.0\columnwidth]{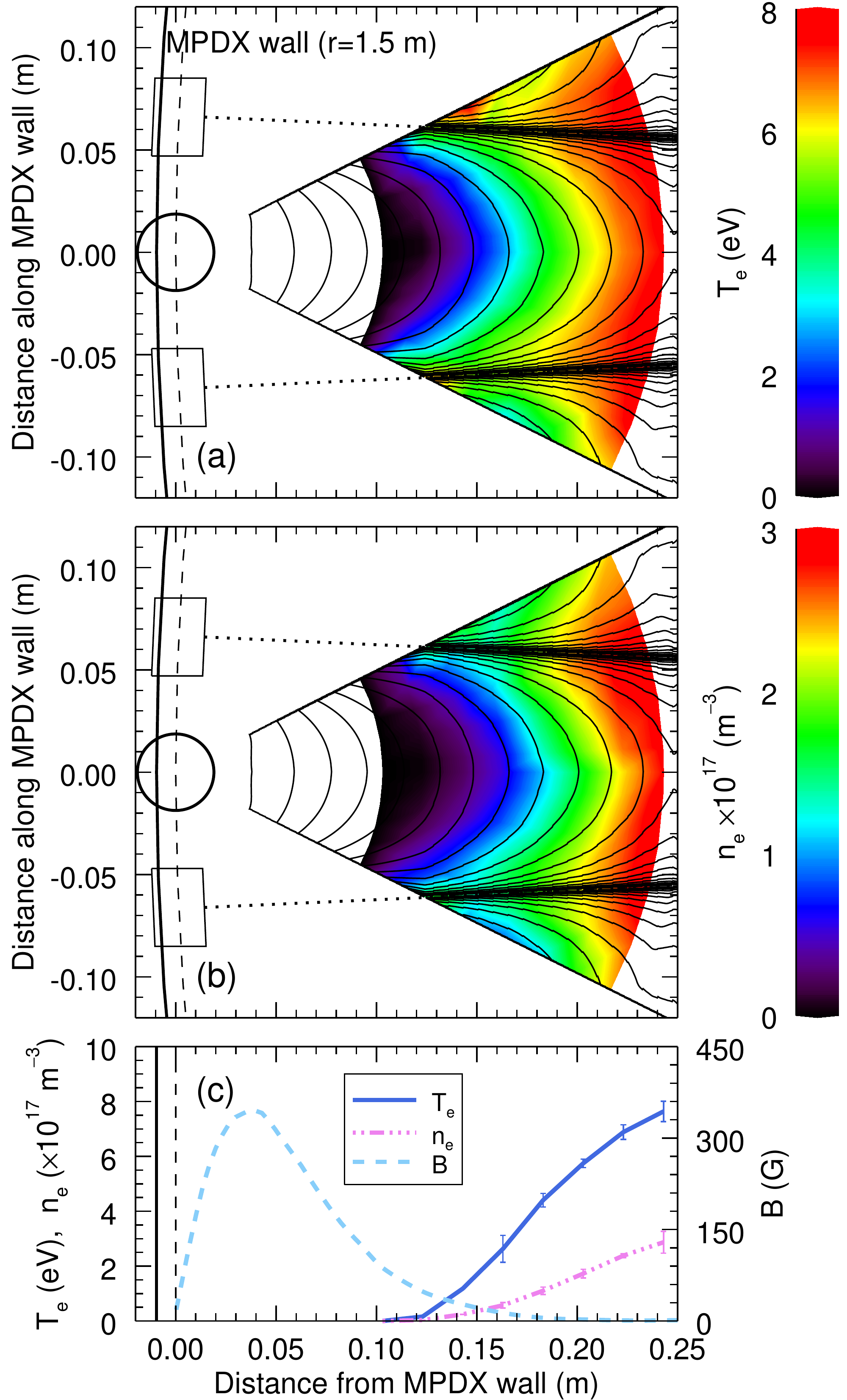}
 \caption{2D plasma profiles of (a) $T_e$ and (b) $n_e$ for a 45 kW discharge in a gas puff corresponding to when $n_n=1\times 10^{18}$ m$^{-3}$ helium.  Isopotentials of the magnetic potential log($\Psi$) found from integrating the measured cusp field are plotted over the data.  A radial cut of the data along the center of the plane is shown in (c) along with the measured value of the magnetic field.  In the unmagnetized core of MPDX, $\beta=100$ for $B=1$ G and $Pm=8$ for $T_i=1$ eV. For v$_\phi=$5 km/s, Re=35, Rm=270 in the MPDX core.}
 \label{fig_radprofile}
\end{figure}

Two dimensional profiles of $T_e$ and $n_e$ of a helium plasma for a 45 kW discharge into a gas puff are shown in Fig. \ref{fig_radprofile}.  The probe was moved in 2 cm radial steps and 6$^o$ toroidal steps over the 54$^o$ sampled (including 0$^o$), creating an 8 point by 11 point plane.  The swept Langmuir probe measured the steady state $T_e$ at each position and the local $n_e$ is determined from the measured ion saturation current, calibrated to the interferometer.  A hall probe was used to measure the 3D magnetic field from the cusp on the same 2D plane as the plasma parameters at a 0.33 cm radial spacing and 3$^o$ toroidal spacing across the 54$^o$.  This magnetic field was integrated to yield the magnetic potential $\Psi$.  Isopotentials of log($\Psi$) are plotted over the data to elucidate how the field lines of the cusp contribute to confinement in MPDX by lowering the effective loss area at the wall, with $\Psi=0$ on the line in the cusp perpendicular to the wall.  The isopotentials are evenly spaced along a chord perpendicular to the wall which indicates the predicted exponential dropoff in magnetic field.  A 1D profile of the plasma along the 0$^o$ line is shown below with the measured magnetic field.  No plasma can be measured between the cusps by the probe until it is 0.14 m from the wall ($r=1.36$ m) where $B_{cusp}$ drops below 40 G.  Prior measurements deeper into the plasma indicate that for $r < 1.32$ m, $T_e$ and $n_e$ are relatively flat, indicating a homogeneous unmagnetized plasma core.  In the core, $\beta=n_ek_BT_e\times 2\mu_o/B^2>>1$, and unity Prandtl number is measured in the experiment, both important for studying the flow driven MHD instabilities.

\begin{figure}
 \includegraphics[width=1.0\columnwidth]{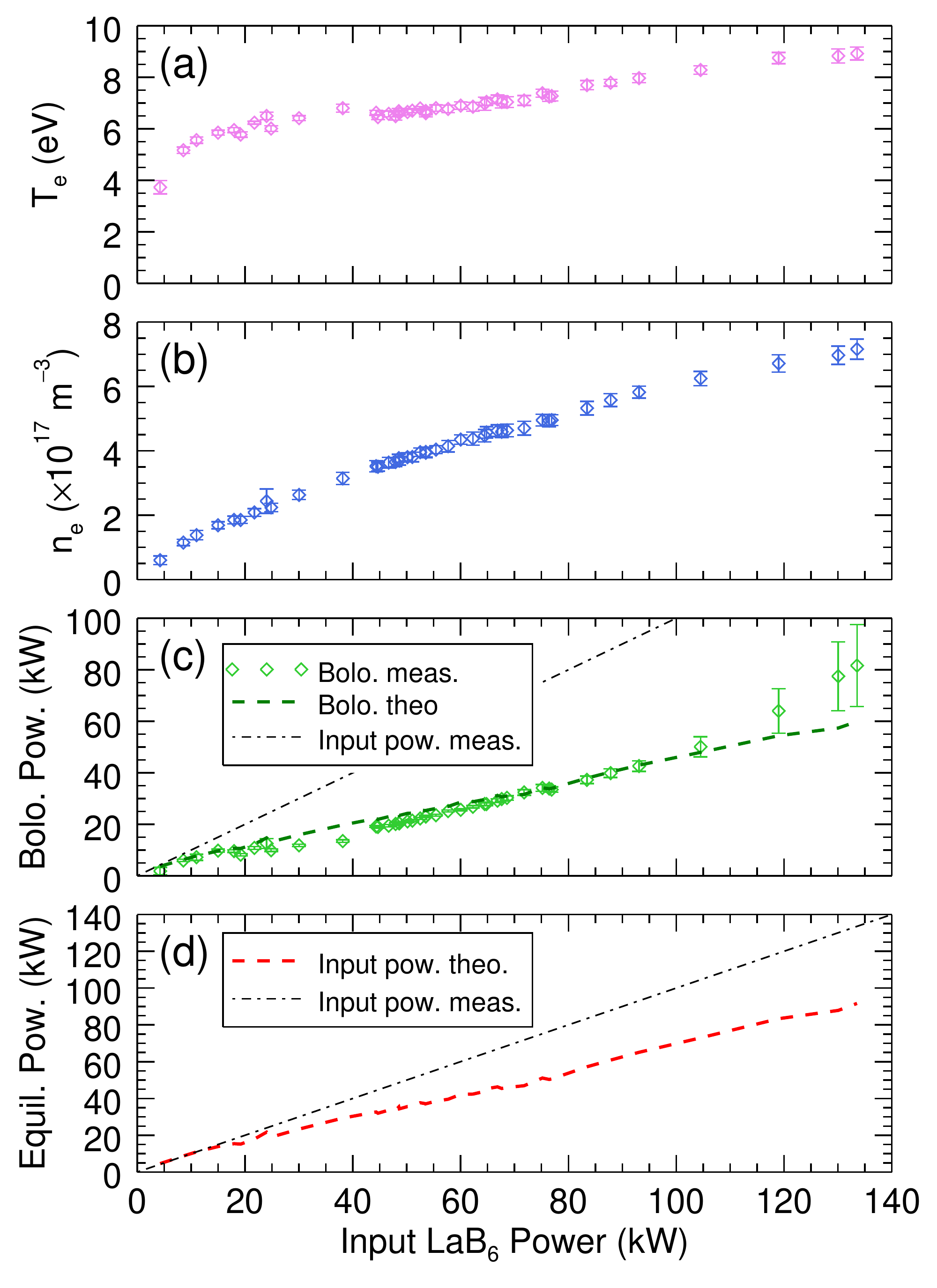}
 \caption{The scaling of (a) $T_e$,  (b) $n_e$ and (c) radiated power measured while varying the input power from 4 - 135 kW into $6.5 \times 10^{-5}$ Torr helium.  The data are used to predict the radiated and equilibrium power required using a confinement a model which is compared to the input power (d) by the LaB$_6$ cathodes.}
 \label{fig_powerscaling}
\end{figure}

The values of $n_e$ and $T_e$ characteristic of the unmagnetized core ($r=1.12$ m) were recorded for a scaling of input power 200 - 450 V, 20 - 300 A (4 - 135 kW) with a constant fill pressure $6.5 \times 10^{-5}$ Torr of helium ($n_n=2\times 10^{18}$ m$^{-3}$), as shown in Fig. \ref{fig_powerscaling}.  The ability to achieve a steady, chamber-filling breakdown with a discharge power as a low as 4 kW is also indicative of a well confined plasma.  The electron temperature saturates at higher input power as the ionization cross section climbs non-linearly and incrementally hotter electrons are required for large changes in density.  Higher $T_e$ should be observed once the ionization percentage becomes sufficiently high.  The total radiated power measured by the bolometer and integrated over the surface of MPDX is shown.  Using the measured the values of $T_e$, $n_e$, and the radiated power, a confinement model can be used to predict the equilibrium power as described in section \ref{sect_sim}, plotted in the final panel.  The experiment will eventually have up to 300 kW of heating power available.

\begin{figure}
 \includegraphics[width=1.0\columnwidth]{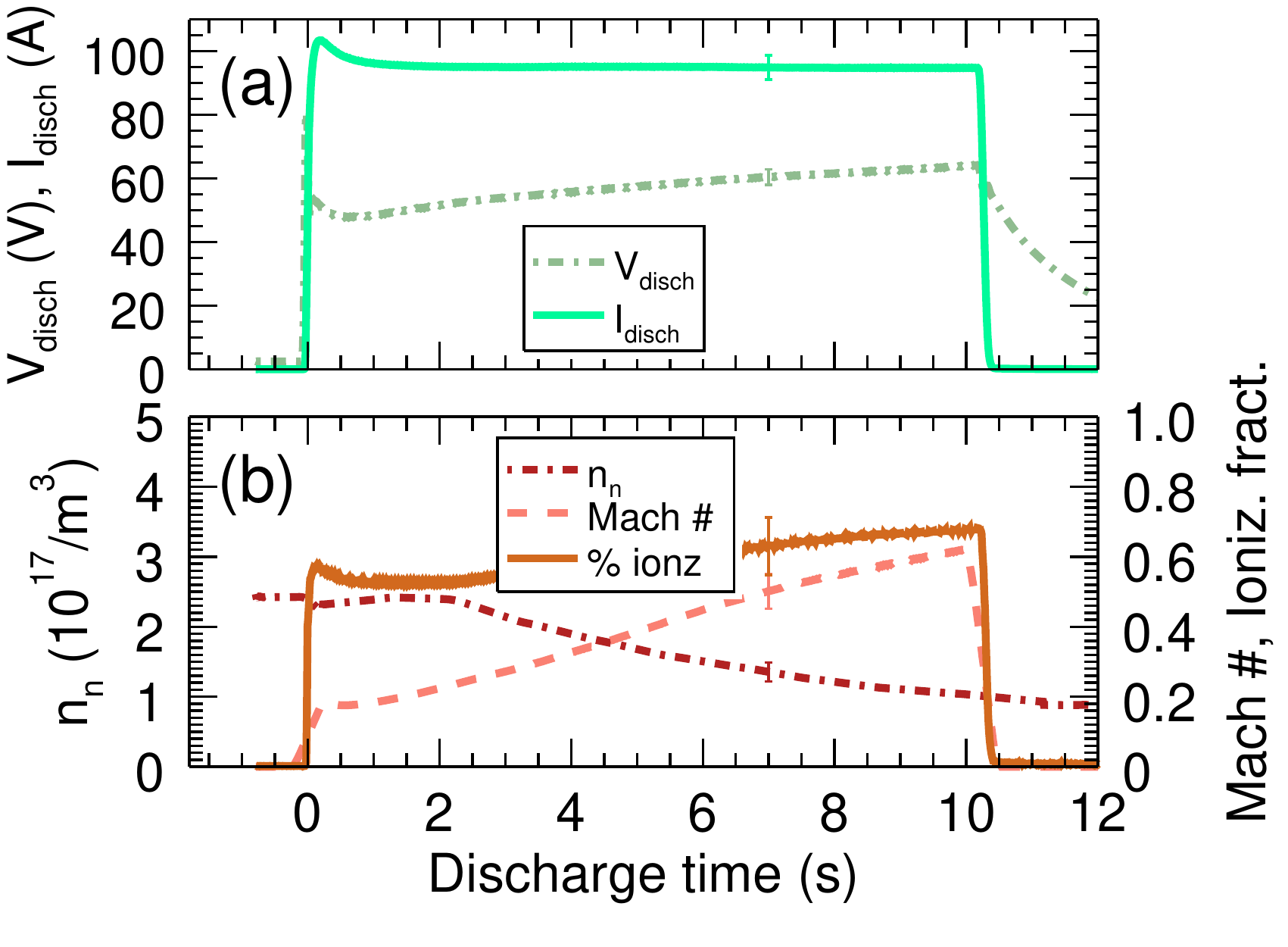}
 \caption{(a) A time trace of the input voltage, current, and power as measured at the anode and cathode and (b) the toroidal flow v$_\phi$ measured by a Mach probe, the neutral density measured by a cold cathode gauge and the ionization percentage.  As the puffed neutral gas is pumped out, the neutral drag decreases and the plasma spins faster.}
 \label{fig_flowtime}
\end{figure}

The second major result is measuring the flow in MPDX.  Only four cathodes and five anodes were used.  The anodes and cathodes were retracted back into the magnetized edge cusp ($r$=1.40 m) at the same toroidal and poloidal locations described above.  In this setup, the only closure path for current is across the cusp field resulting in a ${\bf J}\times{\bf B}$ torque in the toroidal direction.  A two-sided Mach probe was inserted into the plasma oriented to measure the toroidal flow v$_\phi$ in the vicinity of the cathode $r$ = 1.35 m, $\theta=55^o, \phi=45^o$.  An example time trace is shown in Fig. \ref{fig_flowtime}.  To elucidate the flow, the neutral argon gas was puffed into the machine and pumped out over the course of the shot.  In this moderately ionized regime, ion-neutral collisions which dissipate momentum can be comparable to ion-ion collisions which propagate momentum, and the density of the stationary neutral gas plays a role in regulating the flow.  As the neutral density drops, the plasma-neutral gas drag lowers and the plasma rotation velocity for a given torque increases.  In this shot, $T_e=5.0$ eV and the plasma sound speed $c_s=3.5$ km/s, corresponding to a flow speed of v$_\phi$=$2.3$ km/s.  Stirring unmagnetized plasmas with electrodes in a multicusp confinement scheme has been studied in detail in cylindrical device\cite{collins2012}. 

\section{Confinement Modeling of Resistivity and Viscosity}
\label{sect_sim}

In the central unmagnetized region of the MPDX, transport rapidly eliminates any gradients in densities and temperatures allowing for a simple boundary model of plasma confinement.  A zero-dimensional particle and power balance model is used to predict the plasma parameters ($n_e$, $T_e$) in the MPDX given a set of laboratory parameters (fill gas density and input power).  The model includes both plasma physics and atomic physics to predict the particle and energy confinement time.

Volumetric neutral gas ionization by the plasma is balanced by plasma lost by convection to the wall through the magnetic cusps.  Experiments and the best available theory suggested that cusp confinement limits plasma losses by effectively reducing the loss area of the boundary.  The predicted magnetic cusp width $w_c=4\sqrt{\rho_e \rho_i}\sim 0.0008$ m\cite{hershkowitz75_prl,leung76_pf} is proportional to the hybrid gyroradius where $\rho_\alpha=$v$_{th,\alpha} m_\alpha c/eB$ is the gyroradius of a species $\alpha$ based on its thermal velocity v$_{th,\alpha}$, mass $m_\alpha$.  The total length of magnet faces $L_c=$ 220 m in MPDX yielding a convective loss area $A_c=L_c\times w_c=21\times (T_eT_i\mu)^{1/4}B^{-1}$ m$^2$ for the model, which depends on the magnitude of $B$ at the surface of the magnet and weakly depends on $T_e$ and $T_i$.  There are additional losses to the cathodes, anodes, as well as any probes inserted into the plasma denoted as $A_l$.

The volumetric ionization rate $S_p=\left<\sigma_{iz}\text{v}_e\right>n_nn_e$ for an ionization cross section $\sigma_{iz}$ is balanced by a boundary flux $\Gamma_c=0.5n_ec_s$ such that $\iiint S_p \, \mathrm{d} V = \iint\limits_{edge} \Gamma_c \, \mathrm{d} A$.  In the zero-dimensional limit, the volume integral can be replaced by the volume of the machine $\mathbb{V}$ and the surface integral can be replaced with $A_c+A_l$ yielding
\begin{equation}
 \left<\sigma_{iz}\text{v}_e\right>n_en_n\mathbb{V}  = 0.5n_ec_s(A_c+A_l)
 \label{eq_particlebalance}
\end{equation}
Eq. \ref{eq_particlebalance} is transcendental in $T_e$ and independent of $n_e$.  It determines the neutral density $n_n$ in the presence of the plasma in MPDX.  This is related to the experimental fill density $n_f=n_e+n_n$.

The power balance in MPDX is similarly modeled by comparing the power added to the system by LaB$_6$ cathodes and magnetrons $Q_{tot}$ (in kW) to the power lost by volumetric radiation, ionization, and surface particle convection to the walls.  In the 0D limit this becomes
\begin{eqnarray}
 Q_{tot} &=& \left(\gamma_tk_BT_e + E_{iz}\right)0.5n_ec_s(A_c+A_l) + \nonumber \\
   && +  Q_{rad}  +  Q_{\alpha n}
 \label{eq_simpheatbalance}
\end{eqnarray}
The total radiation $Q_{rad}$ arises from electron-neutral collisions $Q_{r,n}=R_{r,n}(T_e)n_en_n\mathbb{V}$ and electron-ion collisions $Q_{r,i}=R_{r,n}(T_e)n_en_i\mathbb{V}$.  The radiation coefficients associated with these processes, $R_{r,n}(T_e)$ and $R_{r,i}(T_e)$, were calculated from atomic rates used in the KPRAD code\cite{Whyte1997KPRAD}.  The heat lost to the neutral gas $Q_{\alpha n}$ consists of electron neutral elastic collisions $Q_{en}=3(m_e/m_i)n_ek_B(T_e-T_n)\nu_{en}\mathbb{V}$, ion neutral elastic collisions $Q_{in}=(3/2)n_ek_B(T_i-T_n)\nu_{in}\mathbb{V}$ and charge exchange $Q_{cx}=n_ek_B(T_i-T_n)\nu_{cx}\mathbb{V}$.  These depend on the elastic neutral gas collision rate for a species $\alpha$, $\nu_{\alpha n}=n_n\sigma_n\sqrt{k_BT_\alpha/m_\alpha}$ with $\sigma_n\sim 5\times 10^{-19}$ m$^2$

The total heat convected to a surface is modeled by the thermal transmission coefficient,\cite{Stangeby1984}
\begin{eqnarray}
 &&\gamma_t = \frac{2}{1-\delta_e}  +  \frac{2}{Z}\frac{T_i}{T_e}  + \nonumber \\
  && -  0.5\ln\left[\left(\frac{2\pi m_e}{m_i}\right)  \left(Z + \frac{T_i}{T_e}\right)  \frac{1}{(1-\delta_e)^2}\right]
 \label{eq_gammat}
\end{eqnarray}
which depends on the secondary electron emission coefficient $\delta_e$ and is only weakly dependent on the species via the mass ratio.  For $T_e<20$ eV, $\delta_e=0$ and $\gamma_t=5-8$.

A given $T_e$ and $n_e$ and $T_i$ uniquely determine neutral fill pressure and input power subject to particle and power conservation.  Eq. \ref{eq_particlebalance} is iteratively solved to find $n_n$.  Eq. \ref{eq_simpheatbalance} is iteratively solved using $T_e$ to calculate $Q_{tot}$.  A more sophisticated model which self consistently calculates $T_i$ based on the ion power balance from collisions and viscous heating, as well as modeling the input beam power as a population of non-maxwellian electrons will be presented in a future publication.  However, the current model provides a reasonable prediction over much of the parameter ranges sought.

The model is benchmarked against MPDX data for $Q_{tot} < 150$ kW in Fig. \ref{fig_powerscaling}.  The measured values of $n_e$ and $T_e$ are used to calculate the radiated power flux.  This rate is compared to the measured bolometer reading to give an approximation for the additional loss area $A_l$.  A value of $A_l$= 0.25 m$^2$ fits the bolometer data best, which closely matches the surface area of the anodes, cathodes and several probes in the machine.  Using this additional loss area, the corresponding values of the equilibrium input power were calculated using Eq. \ref{eq_simpheatbalance}.  The result is plotted as a dashed line in Fig. \ref{fig_powerscaling} (d).  At high power, a fraction of the discharge electron beam is lost directly to the wall before it is thermalized and accounts for the difference between the measured input power (the total beam power) and the predicted equilibrium input power (just the thermalized beam).  This fraction (as high as 25\%) matches predictions based on the theramilzation rate and the loss rate for a 450 eV electron.  This loss fraction must be taken into account when designing the machine to operate at 300 kW coupled to the plasma, and an actual input power from LaB6 may be higher.

\begin{figure}
 \includegraphics[width=1.0\columnwidth]{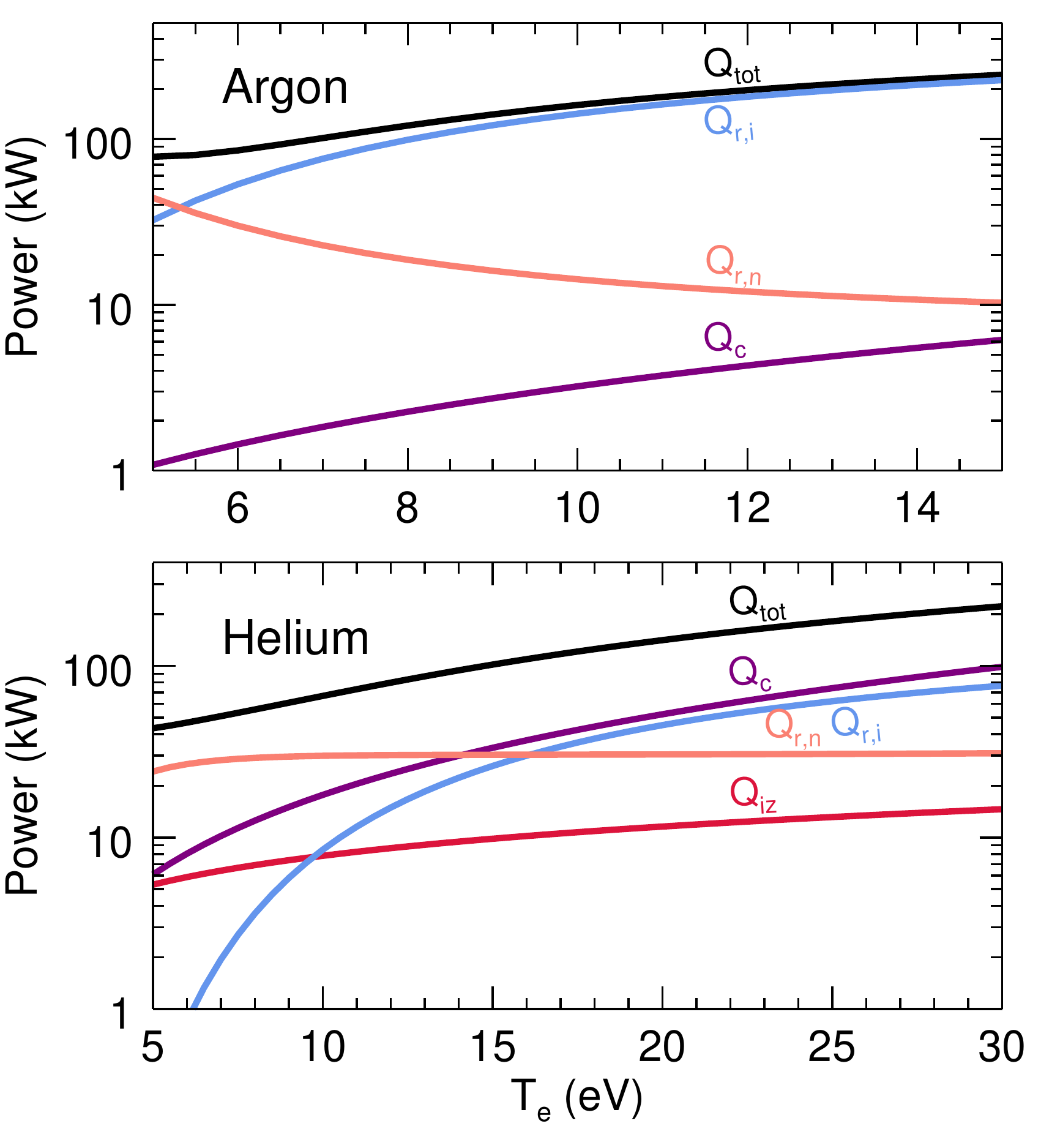}
 \caption{The power balance calculated in the transport model in MPDX for argon ($n_e=2\times 10^{17}$ m$^{-3}$) and helium ($n_e=5\times 10^{17}$ m$^{-3}$) over a scaling of $T_e$.  $Q_{tot}$ is the total power in the experiment, $Q_{iz}$ is the ionization power,  $Q_{c}$ is the convected flux to the wall through the sheath, $Q_{r,n}$ is the radiation from the neutral gas from electron-neutral collisions, $Q_{r,i}$ is the ion radiation from electron-ion collisions and dominates at high $T_e$.  The thermal electron-ion and electron-neutral losses ($<$5 kW) are not shown.  The maximum planned input power to MPDX is 300 kW.}
 \label{fig_powbreakdown}
\end{figure}

The individual power loss mechanisms in the transport model are elucidated in Fig \ref{fig_powbreakdown}.  The total power $Q_{tot}$ is the sum of all the power losses and represents the equilibrium input power required to sustain a plasma with the corresponding plasma parameters $n_e$ and $T_e$.  The power losses are dominated by ion radiation $Q_{r,i}$ in argon due to the high number of emission lines.  While the neutral radiation rate increases with $T_e$, the neutral density decreases from enhanced ionization and there is a net decrease in neutral radiation $Q_{r,n}$ with increasing $T_e$.  The plasma confinement decreases with increasing $T_e$ and the plasma ionization rate and ionization power $Q_{iz}$ increase with increasing $T_e$.  The plasma power convected to the wall through the cusp $Q_{c}$ increases with the higher losses and the higher thermal heat coefficient $\gamma_t$.  This model applies to singly ionized ions only.  When argon doubly ionizes, the radiation rate increases and the model predicts $Q_{tot}$ too low at high values of $T_e$.  However, line radiation ceases for doubly ionized helium and $Q_{tot}$ should be lower than the model predicts at high values of $T_e$.

\begin{figure}
 \includegraphics[width=1.0\columnwidth]{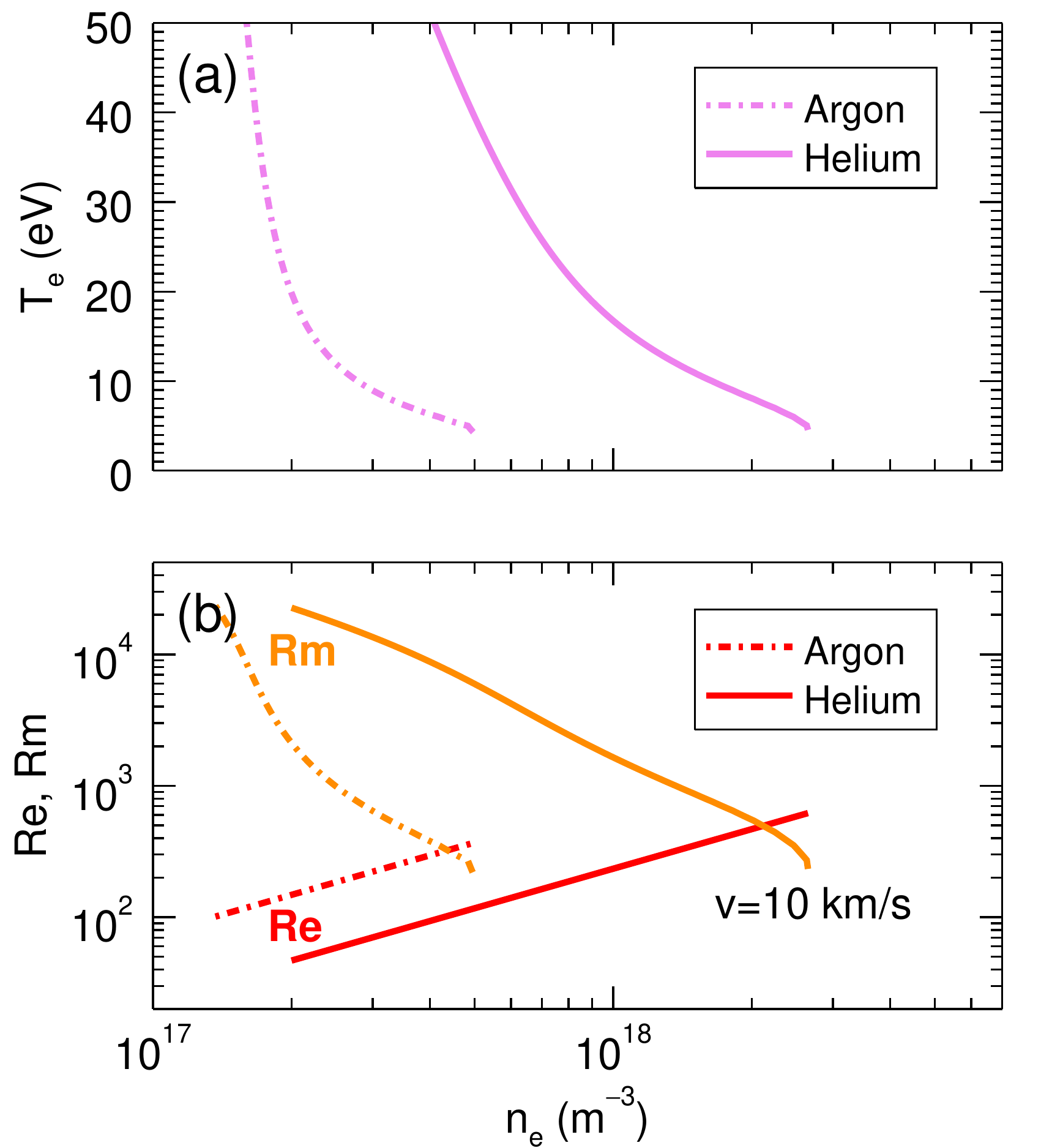}
 \caption{Plasma parameters calculated using the confinement code described above for an input power of 300 kW and v$_\phi=10$ km/s over a scaling of $n_e$ achieved by varying the fill pressure in Argon (dashed) and Helium (solid).  (a) The tradeoff between $T_e$ and $n_e$.  (b) The tradeoff between $Re$ (red) and $Rm$ (orange).  This creates a scaling of magnetic Prandlt number $0.01<Pm=Rm/Re<100$ with the intersections of the lines corresponding to $Pm=1$.}
 \label{fig_rermtradeoff}
\end{figure}

The confinement model predictions are used to calculate $Re$ and $Rm$ plotted in Fig. \ref{fig_rermtradeoff} for v$_\phi=10$ km/s and $T_i=1$ eV, reasonable for the high input power and high ionization fractions.  These predictions address the ability to achieve different types of dynamo action across a scaling of $Pm$ in MPDX.  To generate this plot, the input power is fixed at 300 kW and the plasma density is varied by changing the neutral fill pressure.  This creates a tradeoff between $Rm\propto T_e^{3/2}$ and $Re\propto n_e$.

\begin{figure}
 \includegraphics[width=1.0\columnwidth]{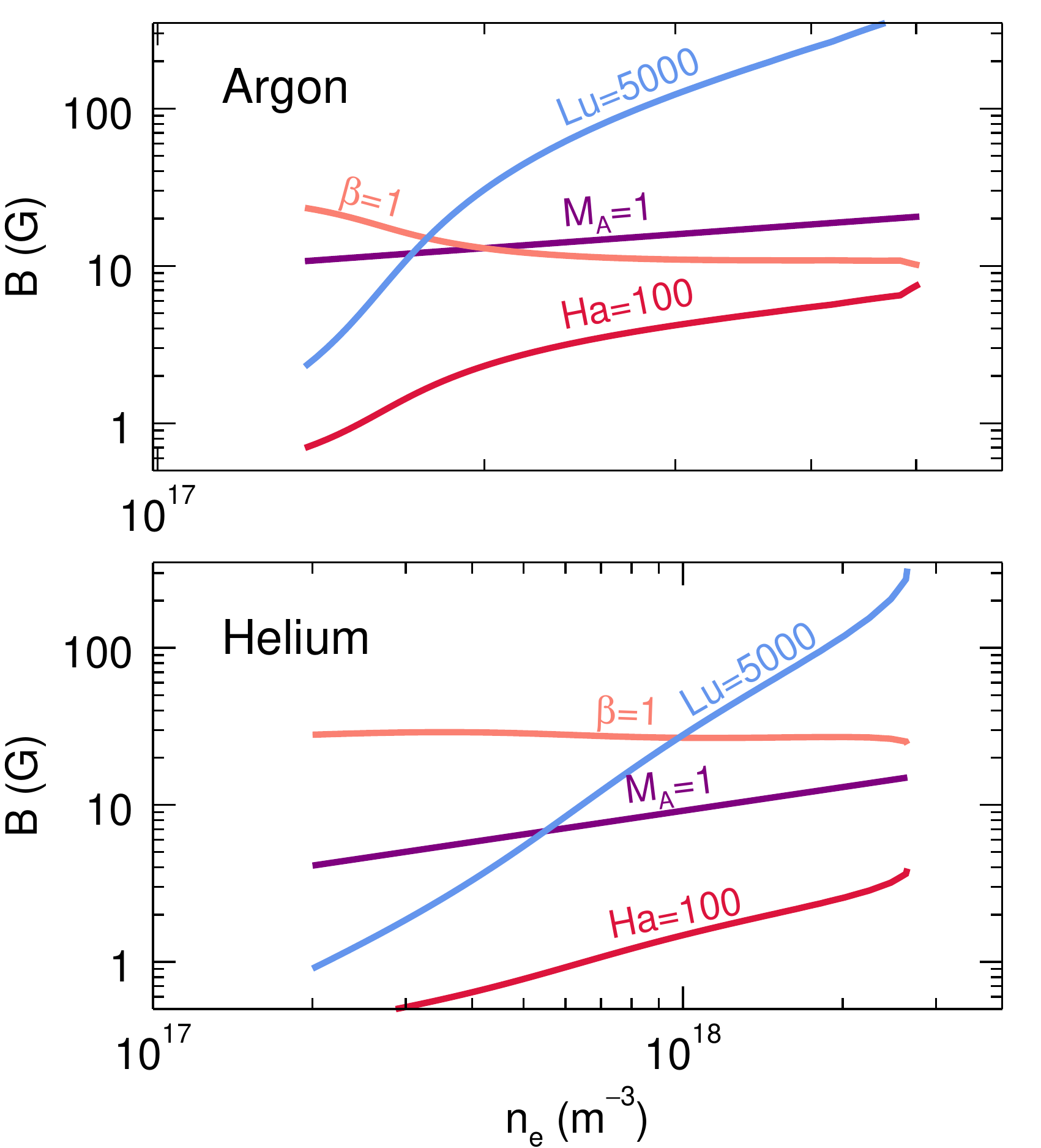}
 \caption{The threshold magnetic field (G) for various astrophysically relevant dimensionless magnetic scaling parameters for (a) argon and (b) helium, the Alfv\'{e}n Mach numbers $M_A=1$, The Lundquist number $S=5000$, and the Hartmann number $Ha=100$.  These values are predicted using the confinement code for a 300 kW discharge for v$_\phi=10$ km/s and 1 G$<B<$100 G.}
 \label{fig_magparamsthreshold}
\end{figure}

The confinement model estimates when magnetic effects become important in MPDX by calculating the magnetic field required for various magnetic dimensionless parameter benchmarks (Fig. \ref{fig_magparamsthreshold}).  The estimates assume v$_\phi=10$ km/s for the same 300 kW discharge described above.  MPDX will be unmagnetized (i.e. a {\it flow-dominated} regime) when the Alfv\'{e}n Mach number $M_A>$ 1 and when $\beta>1$.  This is satisfied for $B<10$ G.

The external Helmholtz coils can be adjusted from 1 G $< B <$ 350 G.  The magnetic effects on the plasma during the transition to magnetically dominated regime, such as the effects on viscosity, can be studied.  The MPDX can also be configured to probe magnetized plasma regimes relevant to laboratory plasma astrophysics.  A Lundquist number $S>5000$ is estimated in MPDX for $B>100$ G.  The Lundquist number is the ratio of the ${\bf J}\times{\bf B}$ torque to the resistive magnetic diffusion force $S=\mu_oL$v$_A/\eta=Rm/M_A$.  The Hartmann number is the ratio of the magnetic force to the viscous force $Ha=BL/\sqrt{\mu\eta}=\sqrt{ReRm}/M_A$ and is important in determining boundary layers.  A Hartmann number $Ha=100$ in MPDX is predicted for $B\sim5$ G.

\subsection{Dynamo Scenarios}
\label{subsect_scenarios}

\begin{figure}
 \includegraphics[width=1.0\columnwidth]{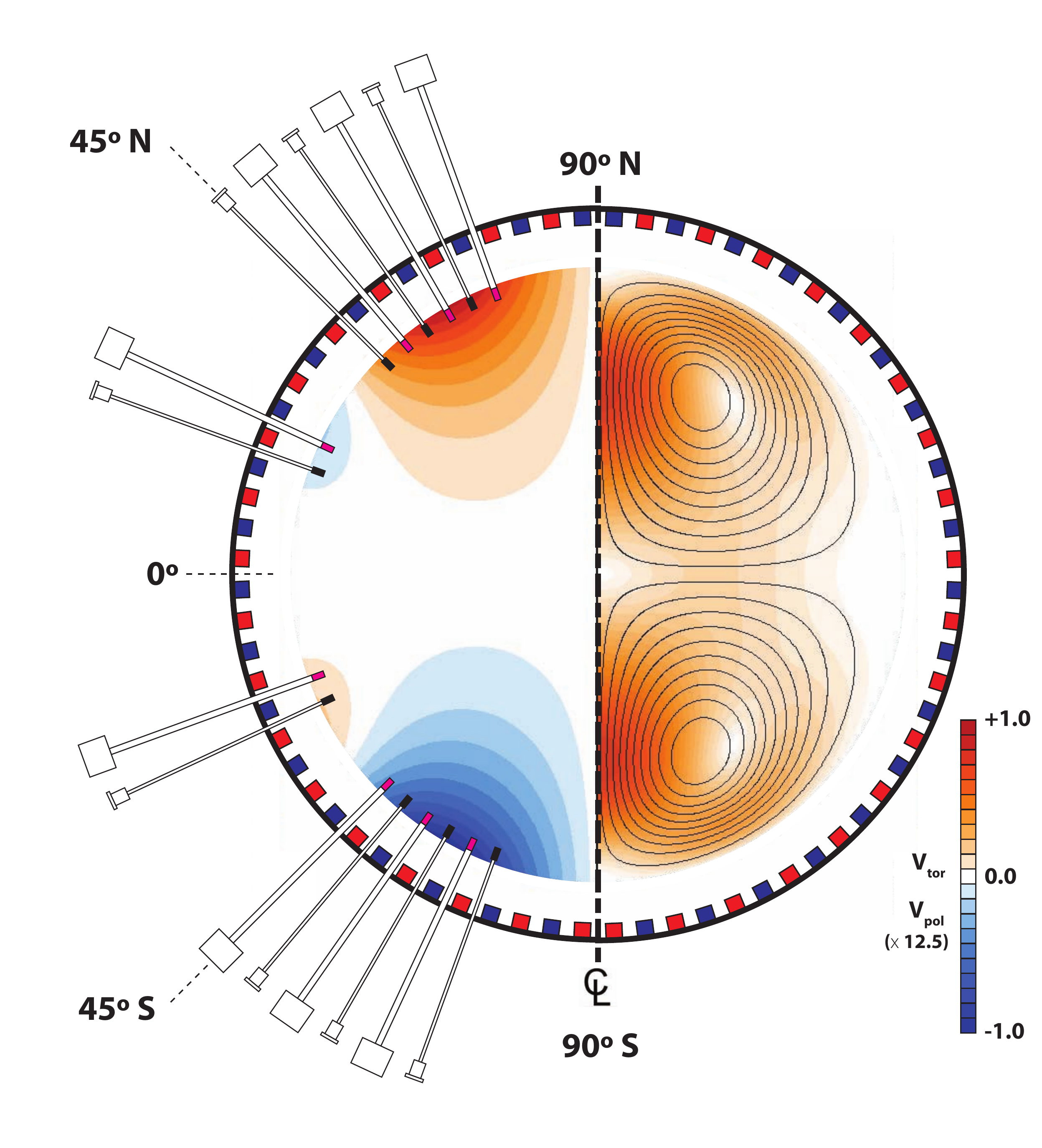}
 \caption{Numerical simulations of the boundary driven flow with  for the flow drive with $Re=300$ in a Von K\'arm\'an geometry with counter rotating flows in each hemisphere.  The poloidal flows (right side) develop self-consistently as a response to the centrifugal forces developing due to toroidal (left) rotation.}
\label{dyn_v1}
\end{figure}

Kinematic dynamo simulations relevant to MPDX have been previously reported \cite{spence09_apj,Khalzov2012, Khalzov2012a,khalzov2013}.  The process for experimental design is briefly reviewed to both emphasize the connection between the dimensionless parameters used in the modeling but also to illustrate how the experiments are planned to operate.  The essence of the stirring is that the axisymmetric multi-cusp magnetic field and cathode stirring allows the azimuthal velocity profile (rotation) v$_\phi(\theta,t)$ to be controlled at the plasma boundary.  Boundary driven flows are modeled by solving the incompressible Navier-Stokes equation 
\begin{equation}
\label{mhd_NS} \frac{\partial\textbf{v}}{\partial t} = \frac{1}{Re}\nabla^2\textbf{v}-(\textbf{v}\cdot\nabla)\textbf{v}-\nabla p,~\nabla\cdot\textbf{v}=0
\end{equation}  
to determine the velocity field subject to the boundary condition v$_\phi(r=a,\theta,t)=f(\theta )$.  
Then, the solution to the Navier-Stokes is tested for dynamo behavior by finding the eigenmodes of the induction equation
\begin{equation}
\label{mhd_b} \frac{\partial\textbf{B}}{\partial t} = \frac{1}{Rm}\nabla^2\textbf{B}+\nabla\times(\textbf{v}\times\textbf{B}),~\nabla\cdot\textbf{B}=0
\end{equation}  
which constitutes a kinematic dynamo problem~\cite{Khalzov2012, Khalzov2012a}.  Note the Lorentz force is not included in Eq.~(\ref{mhd_NS}) when linear dynamo stability is being addressed.  An example of a resulting velocity field is shown in Fig. \ref{dyn_v1} for one of the most promising scenarios for which the transition to a self-excited magnetic field is $Rm_{crit}\sim 300$.

\begin{table}
 \begin{tabular}{|c|c|c|}
 \hline
 \multicolumn{3}{|c|}{ \textbf{Slow Dynamo},  $Re$=150, $Rm$=300} \\
 \hline
 Gas & Argon & Helium  \\
 $n_e$ (m$^{-3}$) & $2\times 10^{17}$ & $1.2\times 10^{18}$  \\
 $T_e$ (eV) & 7.5 & 12  \\
 power (kW) & 100 & 140  \\
 v$_\phi$ (km/s) & 6 & 3  \\
 B$_{eqp}$ (G) & 8 & 3  \\
 \hline
 \end{tabular}
 \vspace{0.0in}
 \caption{List of experimental parameters for achieving slow dynamos in MPDX. }
 \label{tab:MPDX_diagnostics}
\end{table}

Table \ref{tab:MPDX_diagnostics} provides a list of experimental parameters for achieving slow dynamo action based on the simulations described above.  The values of $Re$ and $Rm$ are from predictions of the dynamo growth rate in the simulations.  These uniquely determine $n_e$ and $T_e$ for a given gas ($\mu$), $T_i=1$ eV based on possible flow velocities v$_\phi$.  The dynamo problem is reduced to a confinement problem in MPDX and the power balance code is used to predict the input power and fill density required to achieve the desired plasma parameters.  The equipartition dynamo field that satisfies $m_in_i$v$_\phi^2/2=B_{eqp}^2/2\mu_o$ is used as an approximation for the observable saturation magnetic field for dynamo action.  The predicted magnetic fields are large enough to be measured by Hall probes, as is done in liquid metal dynamo experiments.  The input power required is available, the edge rotation velocities have been measured in similar devices, and the equipartition field is large enough to be easily measured.

\section{Conclusions}
\label{sec:conclusions}

A new experiment for rotating unmagnetized plasma using ${\bf J}\times{\bf B}$ forcing in a magnetic bucket configuration has been described.  The apparatus is suitable for new astrophysically-relevant experiments on flow-driven plasma instabilities.  As in previous magnetic buckets, the arrangement of permanent magnets into a multicusp configuration results in a large volume of uniform, unmagnetized plasma.  A power balance calculation has shown that there is a large range of accessible $Re$, $Rm$ and $Pm$.  

The MPDX continues to upgrade its power systems and diagnostics to approach parameters necessary for exciting dynamo action.  Additional diagnostics to measure the ion distribution function and additional two-dimensional probe drives are planned for use.

\section{Acknowledgments}

This work was funded in part by NSF award no. PHY 0923258, ARRA MRI, NSF award no. PHY 0821899, Center for Magnetic Self Organization in Laboratory and Astrophysical Plasmas, and DOE award DE-SC0008709 Experimental Studies of Plasma Dynamos. C.C. acknowledges support by the ORISE Fusion Energy Sciences Graduate Fellowship.




\end{document}